\documentclass{article}
\usepackage{spconf,amsmath,graphicx}
\usepackage{amsmath,amssymb,amsthm}
\usepackage{xcolor}
\usepackage{algorithm}
\usepackage{caption}
\usepackage{algpseudocode}
\usepackage{booktabs}
\usepackage{float}

\title{Training end-to-end speech-to-text models on mobile phones}
\name{S. Zitha, Raghavendra Rao Suresh, Pooja Rao, T. V. Prabhakar}
\address{Department of Electronic Systems Engineering\\
	Indian Institute of Science, Bengaluru, India, 560012. \\
	Email:\{zithas,poojarao,tvprabs\}@iisc.ac.in,\\
	raghavendrasureshk10@gmail.com}
\begin{document}
%
\maketitle
\begin{abstract}
Training the state-of-the-art speech-to-text (STT) models in mobile devices is challenging due to its limited resources relative to a server environment. In addition, these models are trained on generic datasets that are not exhaustive in capturing user-specific characteristics. Recently, on-device personalization techniques have been making strides in mitigating the problem. Although many current works have already explored the effectiveness of on-device personalization, the majority of their findings are limited to simulation settings or a specific smartphone.  In this paper, we develop and provide a detailed explanation of our framework to train end-to-end models in mobile phones. To make it simple, we considered a model based on connectionist temporal classification (CTC) loss.  We evaluated the framework on various mobile phones from different brands and reported the results. We provide enough evidence that fine-tuning the models and choosing the right hyperparameter values is a trade-off between the lowest WER achievable, training time on-device, and memory consumption. Hence, this is vital for a successful deployment of on-device training onto a resource-limited environment like mobile phones. We use training sets from speakers with different accents and record a 7.6\% decrease in average word error rate (WER). We also report the associated computational cost measurements with respect to time, memory usage, and cpu utilization in mobile phones in real-time.

\end{abstract}
\begin{keywords}
on-device training, personalization, speech recognition, model adaptation, on-device learning
\end{keywords}
\section{INTRODUCTION and MOTIVATION}
\label{sec:intro}
Supervised training of end-to-end speech-to-text (STT) or automatic speech recognition (ASR)  models require a significantly large amount of annotated and transcribed audio data. The commercial STT systems \cite{alexa} are mostly deployed on the centralised cloud with network support infrastructure. With the ubiquitous mobile phone and other embedded devices, there is a demand for reliable and fast loop control edge intelligence based decoding without network infrastructure. In recent times, there is an emergence of distributed training paradigms for STT models that use the rich data collected directly on mobile and other embedded  devices to further train the pre-trained models. Such a method not only reduces privacy risks, but also improves the response time by fine tuning the model parameters.  Furthermore, such systems are constrained by memory and storage ability, decoding speed, availability and reliability of on-device training data and cheap sensor hardware. Hence building end-to-end STT models for such lightweight systems is a challenge in itself. 

STT personalization \cite{speaker_adapt} using acoustic model adaptation aims at fine-tuning the leaner weights of the model to learn the user's voice characteristics such as pitch, accent, and speaking rate in a resource-limited environment like smartphones. In  \cite{Investigation_to_ondevice_personalization}, the authors used a sliding window model to simulate the data consumption as in a mobile environment. The pre-trained recurrent neural network transducer (RNNT) \cite{RNNT1,RNNT2,RNNT3,RNNT4} model weights from the server are quantized to 8-bit integers for deployment. They suggest an approach to minimize memory consumption by splitting the training graph into smaller sub-graphs such that the gradients are calculated separately for the sub-graphs. \cite{ondevicetraining_disorderedspeech} proposes an approach to improve the performance degradation for disordered speech using personalization of the STT models. Both of these studies report the results of a small user study of the personalization strategy on a Pixel 3 phone. However, the manner in which the on-device training was carried out using the quantified model deployed over the phone is not elaborated anywhere.
In \cite{NamedEntities}, the authors investigate the effectiveness of fine-tuning the decoder of the RNN-T model to recognize named entities better. To prevent the personalized model from being indiscriminately accepted, a set of acceptance criteria based on loss and WER on validation data was implemented by \cite{RobustContinuousOndeviceASR}, but the performance of STT on-device training was not reported for both works. The works reported above perform a quantizing and dequantizing of the model weights in between training rounds. Unlike the above discussed approaches, in our work, we persist with the model weights in the quantized form during training, which is desirable for storage efficiency.
Many such approaches focused on reporting acoustic model adaptation in simulation setting, but the deployment constraints of this approach into large scale real-world consumer digital products such as the multiple brands of mobile phone is not reported in entirety. Our objective is to provide a comprehensive overview of the entire on-device training process in mobile phones and conduct experiments on mobile phones instead of system-based simulations.

We make the following contributions in this paper - 
\begin{itemize}
\item We present a methodology for successful implementation of training STT models on multiple brands of mobile phones.
\item We discuss the performance of on-device training with respect to device technical specifications such as memory, Android Version and SoC across a few brands of mobile phones.
\item We present our findings from training the model with various accents on the device over a baseline model to mimic real-life scenarios.
\item We show consolidated results of tuning the training parameters  and their effect on the STT model's accuracy.
\end{itemize}

This paper is organized as follows. We begin by presenting the procedure to train an STT model on-device in section 2, followed by our experimental setup in section 3. We present the comprehensive results from training speech recognition model on mobile phones by varying relevant metrics in section 4.

\begin{table}[t]
\centering
\caption{List of training datasets}
\begin{tabular}[t]{lccc}
\toprule
Dataset&Total Duration&Male&Female\\
&(hrs)&(\%)&(\%)\\
\midrule
LibriSpeech \cite{librispeech}&960&52&48\\
Commonvoice \cite{commonvoice}&2000&45&15\\
TEDlium \cite{tedlium}&250&66&34\\
Fischer \cite{fischer}&2000&53&47\\
\bottomrule
\end{tabular}
\end{table}%

\begin{table}[t]
\centering
\caption{Speaker breakdown for on-device training dataset}
\begin{tabular}[t]{lcc}
\toprule
Training&Male&Female\\
Set&Speakers&Speakers\\ 
\midrule
USA&1&1\\
UK&1&1\\
Australia&1&1\\
India&0&1\\
\bottomrule
\end{tabular}
\label{speaker_breakdown}
\end{table}%


\section{Methodology}
\label{sec:methodology}
\subsection{Model and data} 
We have implemented an end-to-end speech recognition model similar to the DeepSpeech2 \cite{ds2} architecture from scratch in Tensorflow \cite{tensorflow} for our initial deployment of on-device training. The architecture utilizes a well-optimized RNN based training system that does not require extensive pre-processing using phonemes to train.
Our DeepSpeech2 model consists of three convolutional layers (CONV 1-3) for feature extraction, four bi-directional long short term memory (BLSTM 1-4) layers with 1024 units in each direction, and two fully connected layers (FC 1-2) with 1024 units.  
We have trained this model using datasets as mentioned in Table 1. We further augment the dataset by adding colored background noise, Gaussian noise, reverberations to 40\% of the total clean speech dataset at various signal-to-noise ratio (SNR) levels.
The input speech samples are divided into windows of 32 milliseconds with 50\% overlap and converted to frequency domain. From each frame, a 80-dimensional log-melspectrogram is extracted to be given as input to the network. 

One important step in an on-device personalization framework is to adapt the model to user-specific data \cite{domain_adaptation1, domain_adaptation2}. Often, the domain-specific new words used by the user or the user's voice characteristics are unknown to the server-side model. One way to introduce these anomalies to the model in such scenarios is to fine-tune the model with the user's data. Our objective is to re-train the DeepSpeech2 model, and this would require a substantial amount of user-specific data. We created an audio corpus for airplane-cabin specific announcements using a text-to-speech system with nearly 33 voices to simulate different unique speakers with accents. In our current work, we use the dataset of 7 speakers as shown in Table \ref{speaker_breakdown} to train on-device.

\begin{table}[]
\centering
\caption{Total number of trainable parameters for different parts of the model.}
\label{finetuned-models}
\begin{tabular}[t]{lcc}
\toprule
    Layers&Number of trainable &Percentage\\
    & parameters &\\
\midrule
CONV 1-3&45.73k&0.14\\
BLSTM 1-4 &29.11M&96.26\\
FC 1-2 &1.08M&3.6\\
ALL &30.24M&100.0\\

\bottomrule
\end{tabular}
\end{table}%

\subsubsection{Freezing the layers} 
Our DeepSpeech2 model has approximately 30.24M trainable parameters,  and it is not feasible to train this model entirely on a mobile device due to memory constraints. Table \ref{finetuned-models} shows the total number of parameters for different parts of the model. From the Table \ref{finetuned-models}, we can see that 96.26\% of the model parameters comes from the BLSTM 1-4 layers, the FC 1-2 layers contributes only 3.6\% , and the CONV 1-3 layers contributes only 0.14\% of the total parameters. To reduce the memory consumption by the model, we use the model fine-tuning approach, where we try freezing some layers or parts of the model. Hence the output from these frozen layers acts as pre-computed frozen features, and training continues using the rest of the model. This helps to cut down the training time without significantly affecting the accuracy. 

\begin{table}[!t]
\centering
\caption{The quantized tflite models generated from the fine-tuned models and their size in MB}
\label{tflite_sizes}
\begin{tabular}[t]{lcc}
\toprule
    Tflite models&Size (MB)&\#Parameters\\
\midrule
No Frozen &155&30.24M\\
Frozen CONV &154&30.1M\\
Frozen CONV + BLSTM &127&1.08M\\
\bottomrule
\end{tabular}
\end{table}%
\subsection{Implementation} 

Tensorflow Lite toolbox facilitates on-device training by enabling  model size reduction, running faster inference and training. Hence, the large weights of our trained models as defined in section 2.1, are converted to quantized versions in flatBuffer format identified by the .tflite file extension. The inputs and outputs to the TensorFlow Lite model is specified with the help of signature functions to train and infer on-devices.
\subsubsection{CTC loss on phone}
Connectionist Temporal Classification (CTC) \cite{ctc} is an alignment-free, non-autoregressive approach to sequence transduction, used by many decoding frameworks like speech recognition. We use the TensorFlow CTC loss function for on-device training, where it requires the following parameters: labels (ground truth), logits (predicted probability matrix from the trained model), and their respective lengths. 

\begin{table*}[!t]
\centering
\caption{Details about device hardware.}
\label{different_phones}
\begin{tabular}[t]{lcccc}
\toprule
    Device Model&RAM&CPU&OS&SoC (Snapdragon)\\
\midrule
One Plus 7T&8 GB&Octa-core Max 2.96 GHz&Oxygen OS 10 (Android 10)&855 Plus\\
One Plus 5T&6 GB&Octa-core Max 2.45 GHz&Oxygen (Android 10 )&835\\
Redmi Note 7 Pro&6 GB&Octa-core Max 2.02 GHz&MIUI V12.5.1(Android 10)&675\\
Redmi Note 10 Pro Max&8 GB&Octa-core Max 2.3 GHz&MIUI V12.5.2(Android 11)&732G\\
\bottomrule
\end{tabular}
\end{table*}%

\subsubsection{TFlite signatures for on-device training}
Tensorflow in their nightly builds, have come out with new feature called signatures for tflite. These signatures can be manipulated to calculate the loss of the tflite's inference. Furthermore with the loss calculated, we can find the gradients for the models weights, and effectively train the model on device. We use 5 key signatures to facilitate training and inference from the model, namely, Train, Predict, Save, Load and Calculate CTC loss. 

\begin{itemize}
\item Train: implements the back-propagation by calculating the gradient of the loss with respect to the trainable parameters in the model, and update the weights.
\item Predict: performs inference of the model.
\item Save: saves the model's on-device trained weights as a checkpoint file. This allows the weights to be modular and allows us to use them in the future.
\item Load: loads the trained weights from a checkpoint file.
\item Calculate CTC loss: calculates the CTC loss for a given batch of input data.
\end{itemize}

With these signatures we are able to successfully train and personalize our model to our clients. The Train signature later updates the gradients based on the loss calculated. Table \ref{tflite_sizes} shows the size of the quantized tflite models (in Mega Bytes or MB)  generated from fine-tuned models and the 5 signature functions. The `No Frozen' model refers to the tflite model with no frozen layers. The `Frozen CONV' model refers to the ones where we froze all the convolution layers, and the `Frozen CONV + BLSTM' model is generated by freezing all the convolution and BLSTM layers in the baseline model. From the Table \ref{tflite_sizes}, we can deduce that the freezing of all CONV 1-3 layers slightly reduces the total number of trainable parameters.  But freezing the BLSTM layers along with the convolution layers (CONV 1-3 + BLSTM 1-4) reduced the model parameters by 96\%. If none of the layers are frozen, then the whole model is trained across all parameters.

\subsubsection{On-device training}

The training resumes in the mobile phone when the collected number of utterances in the storage cache of the mobile phone is close to a pre-defined number N. We train the quantized model using the train signature for T epochs with a batch size of B. Currently, for our setup, we assume that the transcriptions for the N utterances are available, which is not the case for real-time tasks. After every epoch, we store and restore checkpoints using Save and Load signatures and continue training for T epochs. Upon completion of the training, the stored data will be removed from the cache. The updated model is then used to transcribe newly saved recordings using the Predict signature.

\subsubsection{Stopping criteria}
\label{sec:stopping_criteria}
Training the model continuously on a comparatively small personalized data over a long time can worsen the model due to overfitting. Appropriate stopping criteria should be in place to stop the training and prevent such drifts. In our work, the CTC loss obtained from the signature Calculate CTC loss along with WER is used as the stopping criteria for training. The CTC loss and WER of consecutive epochs are compared. If they are decreasing, the weights of latest epoch are saved and training is resumed. If the metrics increase between two epochs, we continue to train the model for one more epoch. If the metric continues to increase for that epoch, training is stopped. Thus, we have an overhead of one epoch before stopping the training altogether. 

\begin{table*}[t]
	\caption{Time taken per epoch for training the model in system and phone for different batch-sizes}
	\begin{center}
		\begin{tabular}{c c c c c c}
			\toprule
			Batch&Tflite-size&Time per epoch  &RAM &CPU &Time per epoch\\
		 size &in &System-based Simulation &(in GB) &(in $\%$) &Oneplus-7T\\
		 (B)&(MB)&(mins) & & &(mins) \\
			\midrule
               1 &150 &20.96 &2.1  &12  & 76.76                  \\
               2 &151 &13.42 &2.3   &12  &46.92 \\           
               5 &155 &10.9 & 2.9   &12  &19.17\\
               10 &160 &3.96 &3.9  &12 &11.55\\

			\midrule
		\end{tabular}
	\end{center}
	\label{batch_size_vs_epoch_time}
\end{table*}

\begin{figure*}[!t]
    \centering
	\includegraphics[width = 15cm]{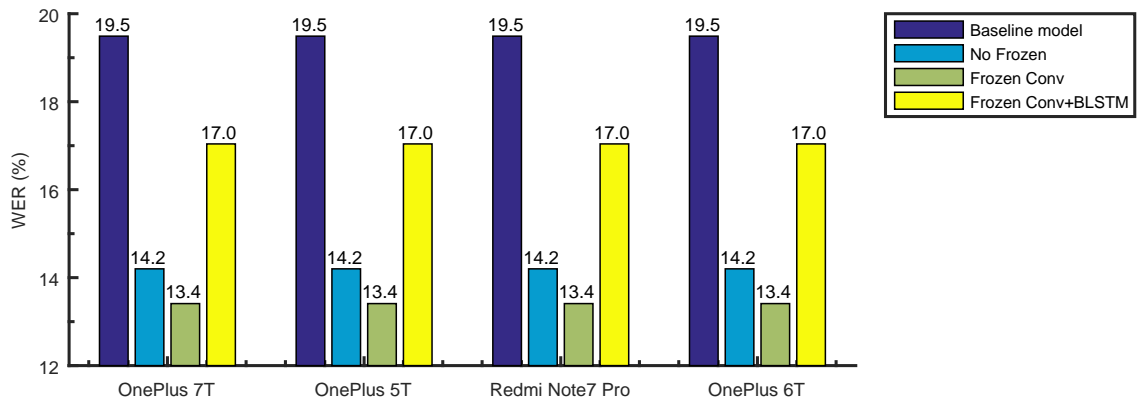}
	\caption{WER after training on different mobile phones with B=5 and learning rate=10e-5}
	\label{wer_vs_phones}
\end{figure*}

\subsection{Information about phone hardware}
In our work, we deploy our on-device training protocol aimed at improving speech recognition across multiple devices from different brands. This provides an insight on deployment constraints such as  training time, accuracy, memory for different hardware specifications. Table \ref{different_phones} shows the hardware specifications of the multiple mobile phones considered in our work.

\section{EXPERIMENTAL SETUP}
\label{sec:experimental setup}
This section presents the approach to investigate the on-device training from dataset generation to profiling memory consumption on phone. It is impractical to obtain thousands of samples for a user to perform on-device personalization. This will require more training time and excess cache memory in a mobile phone environment. Hence, we randomly select 60 speech utterances for training and 10 samples to validate the model, per speaker as defined in Table \ref{speaker_breakdown} . We also limit the average length of the utterances to about 7 seconds and the average label length of these utterances to about 135 characters.

Initially, we study the performance between training in a server and mobile environments. Our intent is to measure the effect of batch size on the size of the tflite model, RAM usage, and speed performance for on-device training and hence, find the best possible configuration for the mobile environment. For this experiment we randomly selected one speaker with US accent and trained the No Frozen model using Adam \cite{adam} optimizer. We set the learning rate of the model to 10e-5, and we implemented the stopping criteria discussed in section \ref{sec:stopping_criteria}  to stop the training and get the best checkpoint. We used the Android Profiler toolkit in Android Studio to access real time data of the memory and CPU utilization of our app. 
Along with batch-size, another critical tuning parameter for training a model on-device is learning rate, as it determines the convergence time and accuracy of results. We selected two learning rates, 10e-5 and 10e-6 to compare. The former is chosen as the Baseline model is trained on-system with same learning rate. We chose another lower learning rate of 10e-6 as we are fine-tuning the model for a particular user. Hence, a lower learning rate would move towards the minimum and prevent overshooting.
Next, we designed an experiment to train using a multiple speaker training set from various regions as mentioned in Table \ref{speaker_breakdown}. We intended to understand the realistic measures of training time and efficacy for various accents in large scale deployment of on-device training. We picked the best performing fine tuned model with the best performing tunable parameters to train. We ran this experiment on a single phone to remove all device related variations.

\begin{table}[!ht]
\centering
\caption{Comparison of different learning rate for fine tuned models on single phone.}
\label{learning_rate_comparison}
\begin{tabular}[ht]{l c c c c}
\toprule
\multicolumn{1}{c}{} & \multicolumn{2}{c}{10e-6} & \multicolumn{2}{c}{10e-5}\\
\midrule
Models&WER&Epochs&WER&Epochs\\
&(\%)&(T)&(\%)&(T)\\ 
\midrule
Baseline &19.49&-&19.49&-\\
\midrule
No Frozen &13.4&12&14.2&1\\
Frozen CONV &13.4&13&13.4&2\\
Frozen CONV+ &&&&\\
BLSTM&19.02&4&17.04&3\\
\bottomrule
\end{tabular}
\end{table}%

\begin{figure*}[!t]
    \centering
	\includegraphics[width = 15cm]{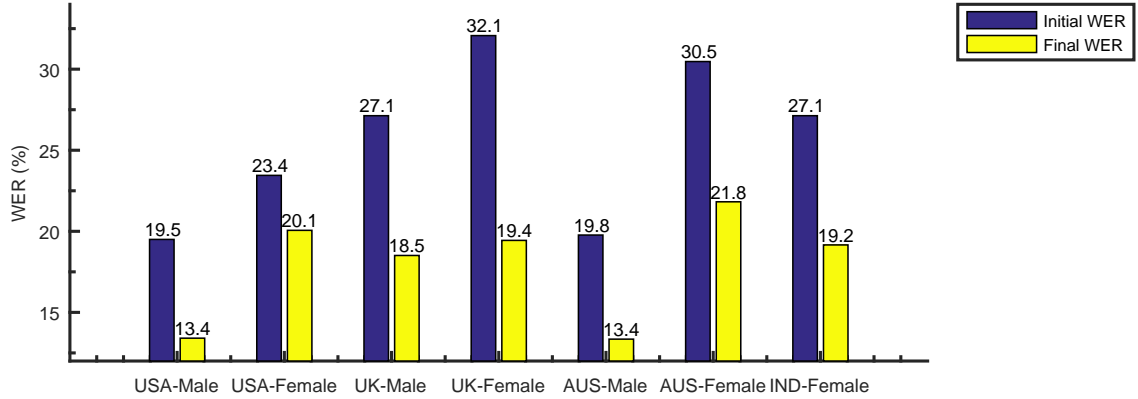}
	\caption{WER for different speakers with B=5, learning rate=10e-5 for Frozen CONV model on OnePlus 7T}
	\label{accents}
\end{figure*}
\section{EXPERIMENTAL RESULTS}
\label{sec:experimental results}
In this section, we present the experimental results from our on-device training settings explained in Section \ref{sec:experimental setup}.
\subsection{Optimal hyperparameters selection}
\subsubsection{Batch size}

Table \ref{batch_size_vs_epoch_time} shows the time taken per epoch during training, the size of the tflite model created for different batch sizes, and the RAM usage during training. The training was carried out on a Oneplus 7T phone with learning rate of 10e-5 and stopping criteria, and the corresponding simulation results are also presented. Based on the results from the mobile phone, we can deduce that the batch sizes affect the size of the tflite model, RAM utilization, and training time. As the batch size is increased from 1 to 10, the size of the tflite model increased from 150 to 160 MB. However, the results clearly indicate that the increase in batch size is significant for training time and RAM utilization. For the batch size of 1, the time taken to complete an epoch while training is nearly 77 minutes, and the memory used is 2.1 GB. As we increased the batch size to 10, the time per epoch decreased to 12 minutes, but the memory consumption increased to 3.9 GB. We, therefore, selected an optimal batch size of 5 for our later experiments.

\subsubsection{Learning rate}

Table \ref{learning_rate_comparison} compares the WERs and the number of training epochs for models mentioned in Table \ref{tflite_sizes} against the Baseline model for both learning rates. The Baseline model achieves a WER of 19.49\%. As expected, the WER of all tflite models for each learning rate is lower than that of the baseline model after training with stopping criteria. Therefore, we can infer that the model has learned from user-specific data for both learning rates. Juxtaposing the results from the two learning rates, we observe that the WER of 10e-6 is 0.8\% lower than 10e-5 for No Frozen model and 2\% higher than 10e-5 for Frozen CONV+BLSTM model. Both learning rates perform the same for the Frozen CONV model. Another deterministic metric is the number of epochs required to converge to the minimum. The No Frozen and Frozen CONV models with 10e-5 learning rate converged within 1 and 2 epoch respectively. In contrast, the same models with 10e-6 learning rate took 10x more epochs to get to similar WER. The training time per epoch is same for both learning rates. Hence, we conclude that the models trained with a learning rate of 10e-5 achieve lower WER faster. In the subsequent experiments, we investigate personalization performance for different phone specification.

\subsection{Experimenting with different phone specifications}
Fig \ref{wer_vs_phones} represents the results from trained models discussed in Table \ref{tflite_sizes} across 4 different phone brands detailed in Table \ref{different_phones}, on a single speaker data. From Fig 1, we infer that Frozen CONV model performs the best in the least time. Frozen CONV+BLSTM model has the least drop in WER despite training for more epochs, meanwhile No Frozen model showed comparable performance to Frozen CONV. The least WER obtained and its respective epoch remains the same for all phones and models. Additionally, we observed that for every phone the CPU utilization is 12\% . Phones with higher RAM like One Plus 7T and Redmi Note 10 Pro Max used 2.9 GB RAM while One Plus 5T and Redmi Note 7 Pro used 2.6 GB, for all tflite models. The most significant influence of phone specification on on-device training was the training time. The training time per epoch for phones with better SoC and CPU specifications, like One Plus 7T and Redmi Note 10 Pro Max, is around 18 minutes, compared to 34 minutes per epoch for devices with lower specifications like One Plus 5T and Redmi Note 7 Pro. 

\begin{table}[!t]
\centering
\caption{Number of epochs to train and training time}
\begin{tabular}[t]{lccc}
\toprule
Training&Number of&Training time\\
~~~~Set&Epoch&per epoch\\ 
&(T)&(mins)\\
\midrule
USA-Male&2&19.23\\
USA-Female&4&19.98\\
UK-Male&7&20.15\\
UK-Female&1&19.95\\
AUS-Male&3&26.55\\
AUS-Female&2&19.92\\
IND-Female&3&19.97\\
\bottomrule
\end{tabular}
\label{accents_time_epoch}
\end{table}%

\subsection{Experimenting with different accents}
So far, we reported the results from a single speaker dataset in all the above experiments. Fig \ref{accents} compares the WER performance before and after training for different speakers from Table \ref{speaker_breakdown} on OnePlus 7T phone. For this experiment, we trained the Frozen CONV model with batch-size, B=5, and learning rate 10e-5 since it was most effective in our previous experiments. The average WER for the different speakers post-training is 17.8\%. Hence, the WER has dropped by an average of 7.6\% across all training sets. This is a substantial improvement over the baseline results. We also observe a bias towards male speakers in the baseline model resulting in lower initial WER. This is expected as percentage of male speakers is higher in the dataset used to train the baseline model. The average WER decreased by 7.03\% for male speakers. While the female speakers have high initial WER, the drop in WER is approximately 9.7\% for all female speakers with the exception of USA-Female. To summarize, on-device training improves the acoustic model results by learning the leaner weights specific to speaker. In Table \ref{accents_time_epoch} we also report the training time per epoch for each speaker and the number of epochs it takes to reach the minimum WER. We see that the number of epochs required to train a model is dependent on the speaker specific characteristics.


\section{CONCLUSIONS}
\label{sec:conclusion}
In this paper, we set out a methodology to  successfully implement  training procedures for end-to-end speech recognition models on mobile phones. Our aim was to provide a comprehensive overview to develop and deploy a complete on-device training framework on the edge devices. We investigated several feasible configurations for the on-device training setup to improve the performance and reduce the complexity in mobile environments. We also conducted extensive experiments to study the impact of system and hardware specifications of mobile phones from different brands on the training procedure. Since the model adaptation relies on user-specific data, we presented our findings from training the model with various accents on the device over a baseline model. Our on-device training results show that there is indeed an improvement in the word error rate over the baseline model on user-specific datasets.

\newpage
\bibliographystyle{IEEEbib}
\bibliography{strings,refs}

\end{document}